\documentclass[aip,graphicx,preprint]{revtex4-1}
\usepackage{graphicx}
\usepackage{dcolumn}
\usepackage{bm}

\usepackage[utf8]{inputenc}
\usepackage[T1]{fontenc}
\usepackage{mathptmx}
\usepackage{etoolbox}

\draft 
\makeatletter
\def\@email#1#2{%
 \endgroup
 \patchcmd{\titleblock@produce}
  {\frontmatter@RRAPformat}
  {\frontmatter@RRAPformat{\produce@RRAP{*#1\href{mailto:#2}{#2}}}\frontmatter@RRAPformat}
  {}{}
}%
\makeatother
\begin{document}


\title{Deep learning with plasma plume image sequences for anomaly detection and prediction of growth kinetics during pulsed laser deposition} 

\author{Sumner B. Harris}
\email[Author to whom correspondence should be addressed: ]{harrissb@ornl.gov}
\author{Christopher M. Rouleau}
\author{Kai Xiao}
\author{Rama K. Vasudevan}
\affiliation{Center for Nanophase Materials Sciences, Oak Ridge National Laboratory, Oak Ridge, Tennessee 37831, United States}
\date{\today}
\begin{abstract}
Materials synthesis platforms that are designed for autonomous experimentation are capable of collecting multimodal diagnostic data that can be utilized for feedback to optimize material properties. Pulsed laser deposition (PLD) is emerging as a viable autonomous synthesis tool, and so the need arises to develop machine learning (ML) techniques that are capable of extracting information from in situ diagnostics. Here, we demonstrate that intensified-CCD image sequences of the plasma plume generated during PLD can be used for anomaly detection and the prediction of thin film growth kinetics. We developed a multi-output (2$+$1)D convolutional neural network regression model that extracts deep features from plume dynamics that not only correlate with the measured chamber pressure and incident laser energy, but more importantly, predict parameters of an auto-catalytic film growth model derived from \textit{in situ} laser reflectivity experiments. Our results are the first demonstration of how ML with in situ plume diagnostics data in PLD can be utilized to maintain deposition conditions in an optimal regime. Further, the predictive capabilities of plume dynamics on the kinetics of film growth or other film properties prior to deposition provides a means for rapid pre-screening of growth conditions for the non-expert, which promises to accelerate materials optimization with PLD.

\vfill
\scriptsize{\noindent Notice: This manuscript has been authored by UT-Battelle, LLC, under Contract No. DE-AC05- 00OR22725 with the U.S. Department of Energy. The United States Government retains and the publisher, by accepting the article for publication, acknowledges that the United States Government retains a non-exclusive, paid-up, irrevocable, world-wide license to publish or reproduce the published form of this manuscript, or allow others to do so, for United States Government purposes. The Department of Energy will provide public access to these results of federally sponsored research in accordance with the DOE Public Access Plan (http://energy.gov/downloads/doe-public-access-plan).}
\end{abstract}
\maketitle 
\section{Introduction}
The recent advent of autonomous synthesis platforms which include multiple in situ or automated diagnostics and characterization techniques \cite{Volk_2023, Burger_2020, Manzano_2020, Coley_2019, Szymanski2023} drives the need to develop machine learning (ML) models that take advantage of this multimodal synthesis data. In situ diagnostics and characterizations are employed in autonomous materials synthesis to derive optimization metrics or correlate growth kinetics with experimental controls\cite{Bianchini_2020, Miura_2021, HarrisACS_2023, Nikolaev_2016}. Additionally, they can be utilized to predict material properties before synthesis and detect anomalous (erroneous) conditions that may arise during long, unsupervised experiments. Pulsed laser deposition (PLD) is a promising physical vapor deposition technique for autonomous synthesis \cite{HarrisAutoPLD} of numerous materials systems due to its compatibility with various optical, electrical, and electron-based diagnostic measurements. Thus, the need to develop ML methods that leverage PLD diagnostic advantages is nascent.

Several types of in situ diagnostics are used during PLD synthesis to monitor either the growing film or the plasma plume: Reflection high-energy electron diffraction (RHEED) \cite{Tischler_2006, Gruenewald_2013}, reflectivity and ellipsometry \cite{Puretzky_2020, Langereis_2006, Gruenewald_2013}, intensified-CCD (ICCD) imaging \cite{Masoud_2009, Lin_2020, Glavin_2014}, and ion/Langmuir probes\cite{Lee_2016, Irimiciuc_2021, Doggett_2009}. Although ML for synthesis has been available for a long time, its applications to in situ diagnostics data is limited. Seminal work was performed by the May \cite{Lee_2000} group wherein neural networks were used with RHEED patterns to predict film properties or forecast RHEED intensity during growth with molecular beam epitaxy. Vasudevan et al. explored the use of other ML algorithms to better understand RHEED image sequences \cite{Vasudevan_2014}, which was then furthered by the Comes \cite{Provence_2020} group. Haotong et. al also recently demonstrated machine learning with RHEED images to construct structural phase maps with respect to PLD synthesis conditions \cite{Liang2022}. Despite the widespread adoption of deep learning across the physical sciences, the adoption in this domain remains highly limited, likely due to sparse or undisciplined datasets. Here, utilizing our autonomous PLD platform, we are able to generate comparatively large amounts of highly-disciplined PLD synthesis data that enables us to explore new ML methods with in situ diagnostics for thin film growth.

Plume imaging data from ICCD cameras (or even just visual inspection or common CMOS camera images) has promise to increase the reproducibility of PLD experiments or be used to predict characteristics of the film such as growth rate, domain size, crystallinity, or defect density which alter material properties. Key features of the PLD plume and its expansion dynamics such as brightness and color (optical emission)\cite{Glavin_2015}, angular distribution, maximum kinetic energy\cite{Lin_2020}, the presence of "fast" or "slow" components \cite{Geohegan_1995, Schou_2007} can be correlated to properties of the thin film or be indicators that the plume conditions are consistent among experiments. Moreover, if plume dynamics alone can act as a reasonable predictor of film properties after synthesis, quick ICCD imaging experiments can be used as a low fidelity surrogate\cite{Kuya_2011, Tran_2020} to more rapidly explore the synthesis parameter space and reduce time/resources wasted on full growth and characterization with conditions that would produce poor quality outcomes. However, analysis of plume image sequences can be a time consuming - i.e., an "offline" effort - and is likely a major contributing factor to the lack of widespread adoption as an on-the-fly, quantitative diagnostic for synthesis. Also, it is generally unknown how certain aspects of the plume are correlated to materials properties which is another major barrier for utilizing ICCD imaging in real-time. Thus, ML with plume images may provide a method of extracting key features of the plume and correlating them with quantities of interest. 

Here, we seek to prove the viability of deep learning with ICCD images for monitoring plume conditions or predicting film properties in real-time during PLD synthesis. We use a $(2+1)$D convolutional neural network (CNN) to extract deep features from ICCD image sequences to correlate plume dynamics to PLD conditions and film growth kinetics. To detect anomalies in the anticipated conditions during unsupervised autonomous synthesis routines, we show that ICCD image sequences alone can predict the chamber pressure and laser energy - two typically measured qualities - but more importantly, predict and improve the accuracy of parameters of a thin film growth model. Our results demonstrate the potential of routine plume imaging for effective anomaly detection and prediction of growth kinetics or other film properties that can be used to improve and accelerate PLD synthesis experiments.
\section{Results and Discussion}
\subsection{\label{datagen}Data Generation}
The ICCD image sequences, growth parameters, and growth kinetics parameters were generated during an autonomous PLD synthesis campaign which is separate from this work. Complete details of the experimental results, PLD chamber design, and diagnostic methods can be found elsewhere \cite{HarrisAutoPLD, HarrisACS_2023}. Briefly, 127 WSe$_2$ films were grown to approximately 1 monolayer thickness on 90 nm SiO$_2$/Si substrates using PLD by co-ablating two targets (WSe$_2$ and Se). Each sample had a combination of background pressure \textit{P}, substrate temperature \textit{T}, laser energy on the WSe$_2$ target \textit{E}$_1$, and laser energy on the Se target \textit{E}$_2$. During growth, a sequence of 50 ICCD images was taken with delay times (relative to the excimer laser pulse) from 2-150 $\mu$s and, simultaneously, laser reflectivity was used to measure the reflected contrast change vs. time of the sample to detect sub-monolayer nucleation and growth of the WSe$_2$ film. Therefore, each individual deposition has a corresponding sequence of 50 ICCD images, a set of (\textit{P}, \textit{T}, \textit{E}$_1$, \textit{E}$_2$), and a laser reflectivity curve.

Each ICCD image is 1 channel (16-bit) with $1024\times1024$ pixels. The images are Gaussian filtered to reduce noise, resized to $40\times40$ pixels, and the log of the intensity is taken to enhance weak characteristics in each image (because the luminous intensity of the plume spatiotemporally varies by orders of magnitude). This gives each image sequence a depth, height, and width $D\times H\times W = 50\times40\times40$. Lastly, the laser reflectivity curves are fit with an auto-catalytic growth model that can represent the nucleation and growth of the film in terms of fractional monolayer coverage \cite{Puretzky_2020}. The three growth kinetics parameters returned by this model are denoted $s_0$ and $s_1$ for the sticking coefficients of plume species to the substrate and film, respectively, and $J$ which is the flux of arriving species. For the purposes of this work, we do not claim that this is the correct growth model or mechanism but rather use it as a fitting equation that effectively reproduces the observed laser reflectivity curves so that the growth kinetics curves can be predicted by deep learning.
\subsection{Machine learning models and training}
\begin{figure*}
\includegraphics{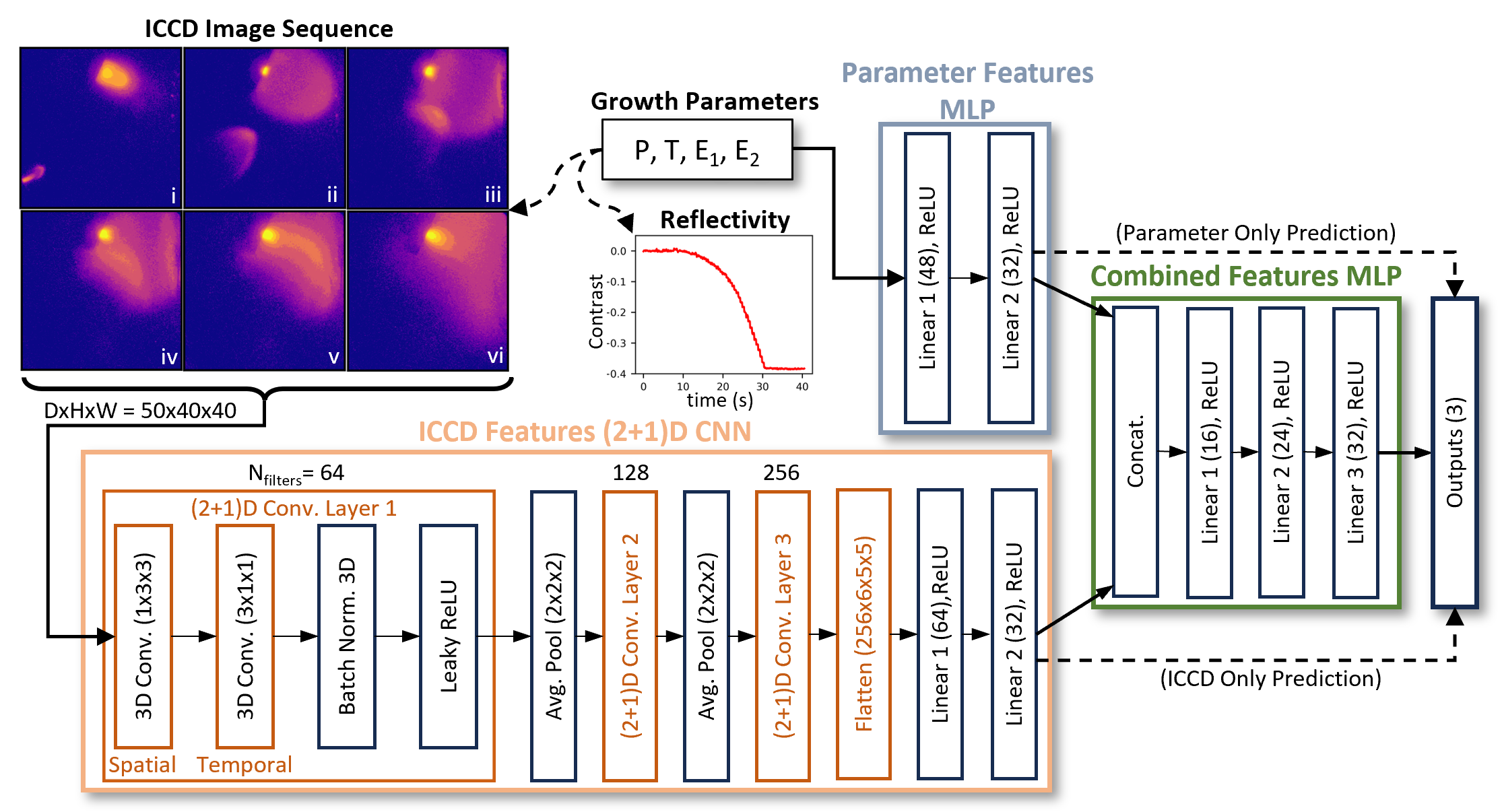}
\caption{\label{network_diagram}Schematic diagram of the neural networks used for multi-output regression with intensified-CCD (ICCD) image sequences and/or growth parameters as inputs. Pulsed laser deposition (PLD) growth for each sample is carried out by selecting a chamber pressure, substrate temperature, and laser energy on each PLD target (P, T, E$_1$, E$_2$). During the growth, an ICCD image sequence and in situ laser reflectivity curve are collected. The image sequence is fed into a $(2+1)$D convolutional neural network (CNN) to extract deep features from the plume dynamics. For experimental anomaly detection, P, E$_1$, E$_2$ are predicted from the ICCD features alone. The growth kinetics parameters (s$_0$, s$_1$, J) derived from reflectivity measurements are predicted by either a multilayer perceptron (MLP) using the growth parameters, the $(2+1)$D CNN, or a by combining the features from both in a final MLP.}
\end{figure*}
Fig. \ref{network_diagram} shows a schematic of the ML networks used for this this study. We used a $(2+1)$D CNN\cite{Tran_2018} to extract deep features from the ICCD image sequences for multi-ouput regression. A $(2+1)$D convolution splits a 3D convolution into successive 2D spatial and 1D temporal convolutions with the benefit of a reduced number of parameters and increased nonlinearities relative to a standard 3D convolution\cite{Tran_2018}. To make predictions from (\textit{P}, \textit{T}, \textit{E}$_1$, \textit{E}$_2$) alone for comparison, we use a simple multilayer perceptron (MLP). For a mixed input model which includes both ICCD images and (\textit{P}, \textit{T}, \textit{E}$_1$, \textit{E}$_2$), we combine the output from the ICCD and growth parameter MLP into a final, combined MLP.

The $(2+1)$D convolution layers consist of 2 sequential 3D convolutions with kernel sizes of ($1\times3\times3$) and ($3\times1\times1$) for the spatial and temporal components, respectively, followed by a 3D batch normalization layer and a leaky ReLU activation \cite{Maas_2013}. The ICCD features are extracted using 3 successive $(2+1)$D layers (with 64, 128, and 256 filters) with an average pooling layer after each with a ($2\times2\times2$) kernel to downsample the data. The output of the last $(2+1)$D layer is flattened into a $256\times6\times5\times5$ tensor and passed through 2 final fully connected layers (Linear 1, Linear 2) with 64 and 32 nodes each. Features are extracted from the growth parameters by feeding the 4 inputs into a simple MLP, 2 fully connected layers with 48 and 32 nodes respectively. For predictions of \textit{P}, \textit{E}$_1$, and \textit{E}$_2$ using only ICCD images, the Linear 2 layer is reduced to 3 outputs for regression and the same is done for predictions of $s_0$, $s_1$, and $J$ with the growth parameters MLP. Finally, the mixed input model for predictions of growth kinetics concatenates the outputs of each sub-network and feeds through a 3 layer MLP with 16, 24, and 32 nodes in each layer.

The 127 sample dataset for this study can be considered quite small in the context of typical deep learning studies. In the context of materials synthesis with physical vapor deposition techniques, this is a typical size compared to other similar autonomous synthesis studies, typically numbering a few 10s to 100s of samples \cite{Shimizu_2020, MacLeod_2020, Chang_2020}. During training, the data is shuffled and split 70/30 into training and validation sets, 88 and 39 samples respectively, and training is done with batch gradient decent. To increase the variety in this small dataset and improve model generalization, we augment both the image and growth parameter data. The image sequences are transformed with random rotations, translations, shear, and scale in each epoch such that the same transformation is applied only in the spatial dimensions for each temporal frame. Similarly, \textit{P}, \textit{T}, \textit{E$_1$}, \textit{E$_2$} are augmented by adding Gaussian noise with an appropriate variance to simulate the measurement accuracy of \textit{P} and \textit{T} and natural shot-to-shot stability of the excimer laser. We found that training with data augmentation improved model predictions. We used the mean square error (MSE) loss function and the Adam \cite{Adam} optimizer. Hyperparameter tuning was done using Ray Tune \cite{liaw2018tune} with the Optuna alogrithm \cite{optuna_2019} and asynchronous successive halving (ASHA) \cite{LiJRGBHRT20} for early-stopping of poor performing runs. The hyperparameters that were optimized were the Adam learning rate, L2 regularization, and number of nodes in the linear layers. Hyperparameter tuning was done for each model discussed in this paper.
\subsection{\label{anomsec}Anomaly Detection}
In PLD, several usually unmonitored factors can affect the reproducibility between depositions. One well known issue in PLD is that material is deposited on the chamber's laser window over time which attenuates the laser energy and changes the plume dynamics \cite{Cazzaniga_2017, Ohnishi_2005}.
Another common unknown is current state of the PLD target where changes during ablation or between similar targets can cause a variation in plume dynamics and film stoichiometry \cite{Cesaria_2021}.
Because all of these effects cause changes in the plume dynamics, deep learning can be used to encode the essential features of the plume dynamics under different \textit{P} and \textit{E} for a single experimental campaign and be used for feedback in future experiments. For example, when attempting to replicate depositions that are separated by significant amounts of time or other interim experiments - as would be the case in a User Facility, for example - an ICCD image sequence can be fed to a model to compare its predictions to the current conditions. If the model consistently predicts parameters that are far from current values, this is a clear indicator of anomalous, unreplicated plume conditions. Moreover, during a rapid sequence of depositions like in an autonomous synthesis campaign, model predictions can be used for compensatory purposes, such as mitigating the effects of laser entry window coating or resolving problems in the process gas flow control system. 
\begin{figure}
\includegraphics[width=\linewidth]{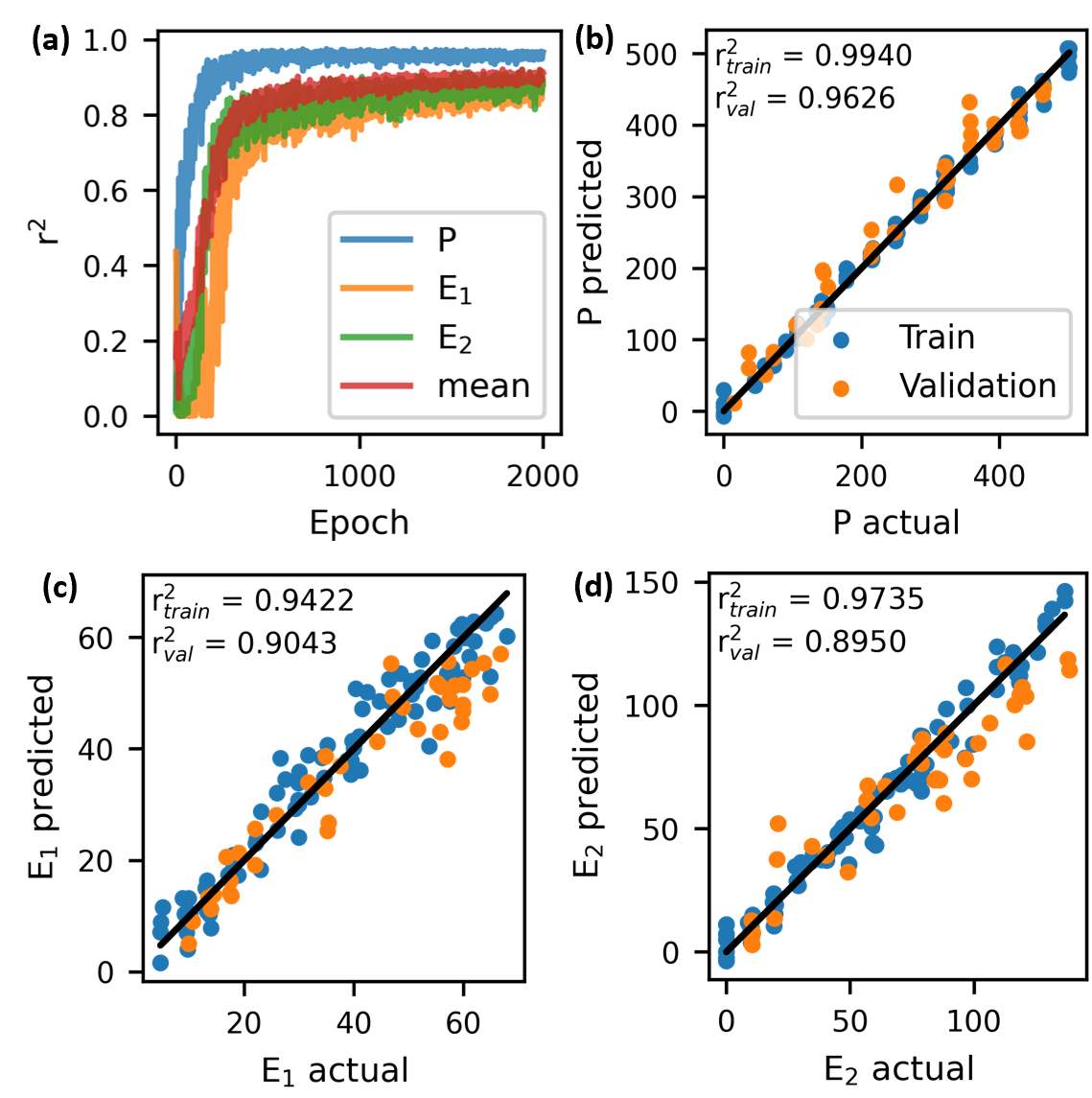}
\caption{\label{anom_detect}Training results for the anomaly detection model which predicts the chamber pressure \textit{P} and laser energies \textit{E}$_1$ and \textit{E}$_2$ from intensified-CCD image sequences using a $(2+1)$D convolutional neural network. (a) Validation set coefficient of determination r$^2$ vs. epoch for each predicted quantity and the mean r$^2$. Predicted vs actual (b) \textit{P}, (c) \textit{E}$_1$, and (d) \textit{E}$_2$ with a mean r$^2$ = 0.921 on the validation set.}
\end{figure}

For anomaly detection, we use only the ICCD image sequence and train the model to predict \textit{P}, \textit{E}$_1$, and \textit{E}$_2$ (ICCD Features $(2+1)$D CNN in Fig. \ref{network_diagram}). The plume dynamics within the 2-150 $\mu$s delay time showed no correlation with the substrate temperature \textit{T}, as anticipated. Fig \ref{anom_detect}a shows the coefficient of determination r$^2$ for \textit{P}, \textit{E}$_1$, \textit{E}$_2$, and the mean of all three as a function of training epoch. The model was trained for 2000 epochs and a checkpoint was taken at the epoch with the highest mean r$^2$. The pressure prediction has the greatest performance with r$^2_{val}$ = 0.963 on the validation set with \textit{E}$_1$ and \textit{E}$_2$ having r$^2_{val}$ = 0.904 and r$^2_{val}$ = 0.895, respectively, giving an average r$^2_{val}$ = 0.921. For comparison, we also trained a model using only a single ICCD image (2 $\mu$s delay) rather than the full sequence, using the same model architecture but without the temporal convolution. This single image model performed significantly worse than the image sequence, with r$^2_{val}$ = 0.850, 0.814, and 0.810 for \textit{P}, \textit{E}$_1$, and \textit{E}$_2$, respectively, a mean r$^2_{val}$ = 0.825. These results indicate that deep learning with ICCD image sequences can be used to effectively predict PLD processing parameters for anomaly detection and that including both the spatial and temporal components of the plume dynamics in deep learning greatly increases model accuracy.
\subsection{Prediction of growth kinetics}
Prediction of film properties or growth kinetics from plume diagnostics is attractive because it enables partial exploration of the PLD parameter space with a lower time and material cost, which accelerates the rate of discovery or optimization of materials. While some film properties may be dominated by thermodynamics, features of the plume may be able to explain some portion of the variance within a dataset. Different species within a multi-element PLD plume (e.g. WSe$_2$) take on different angular distributions which can affect film stoichiometry \cite{Ojeda-G-P_2018} thus altering film properties. Depending on the laser and pressure conditions, the plume is composed of a mix of atomic and molecular species, nanoparticles, and large particulates \cite{Geohegan_1998, Kroto_1985}. These "components" affect the plume expansion dynamics and optical emission which can be detected by ICCD imaging. In the context of autonomous synthesis with PLD, ICCD image sequences could be used to provide additional information to the ML models that determine the sequence of experiments. For instance, if ICCD data can be encoded to predict at least some of the variance in an optimization variable, it shows promise to be utilized for deep kernel learning \cite{Liu_2022}.
\begin{figure*}
\includegraphics[width=\linewidth]{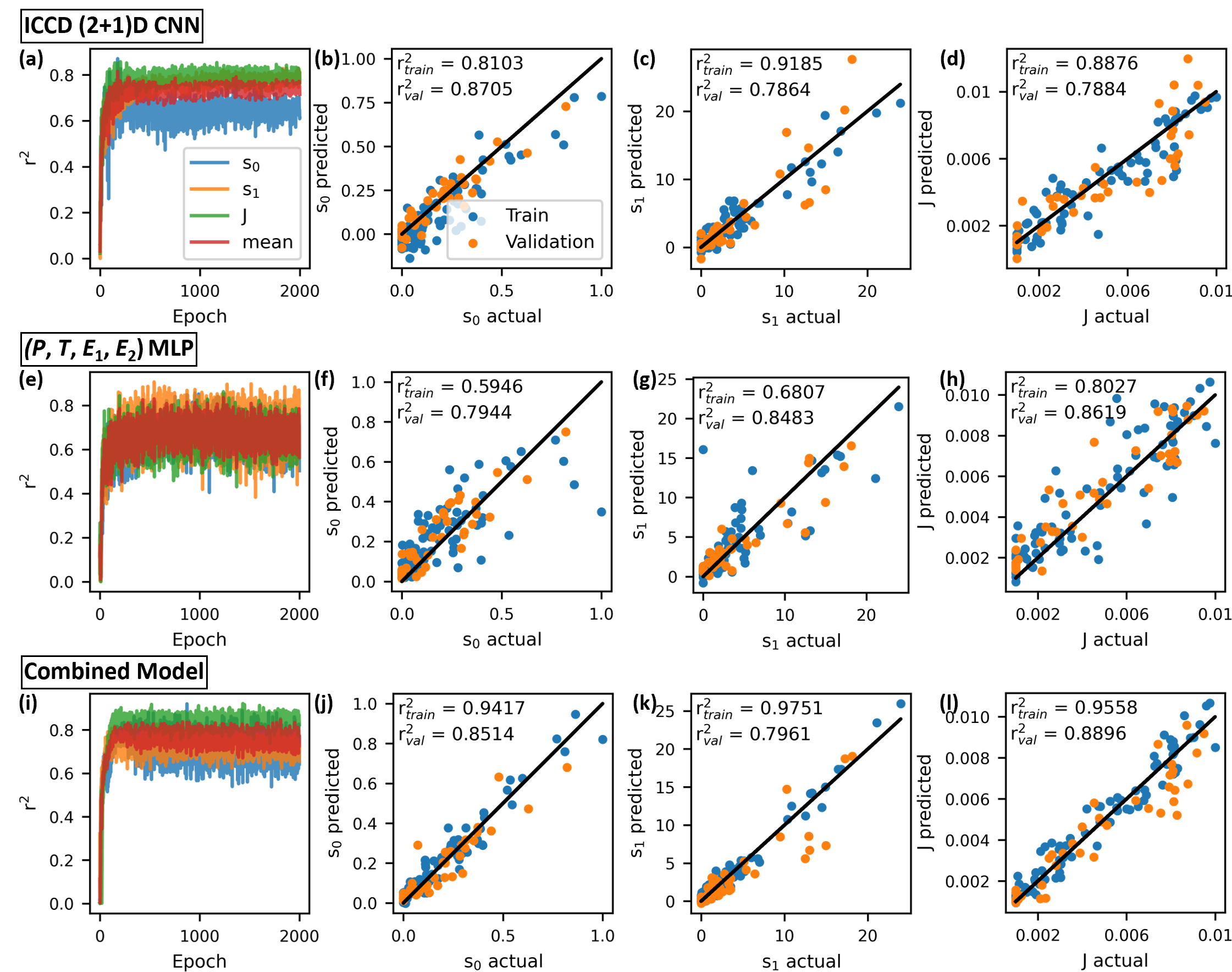}
\caption{\label{LR_predict}Training results for the WSe$_2$ growth kinetics models which predict the kinetic parameters $s_0$, $s_1$, and $J$ using either a $(2+1)$D convolutional neural network (CNN) with intensified-CCD (ICCD) image sequences, a multilayer perceptron (MLP) with the pressure, substrate temperature, and laser energies (\textit{P}, \textit{T}, \textit{E}$_1$, \textit{E}$_2$), or both combined. (a)-(d) ICCD $(2+1)$D CNN model results have a mean r$^2_{val}$ = 0.815. (e)-(h) Growth parameter MLP results with a mean r$^2_{val}$ = 0.835. (i)-(l) Combined model using both ICCD and growth parameter features performs the best with mean r$^2_{val}$ = 0.847.}
\end{figure*}

To demonstrate the predictive capabilities of ICCD imaging on quantities related to the growing film, we train several models to predict the growth kinetics of the WSe$_2$ films as determined from laser reflectivity. An example reflectivity curve is shown in Fig. \ref{network_diagram} and the 3 target quantities are $s_0$, $s_1$, and $J$ (described in Sec. \ref{datagen}). We trained three different models to compare performance between ICCD features only, growth parameter (\textit{P}, \textit{T}, \textit{E}$_1$, \textit{E}$_2$) features only, and both features combined (Fig \ref{network_diagram}). Each model was trained for 2000 epochs and the checkpoint with the highest mean r$^2_{val}$ was saved. For the combined model, the optimized model states from the ICCD $(2+1)$D CNN and the growth parameter MLP were loaded and frozen so that only the combined MLP (see Fig.\ref{network_diagram}) was trained and optimized. Unfreezing the ICCD and growth parameter feature sub networks and retraining did not lead to significantly different results. Fig. \ref{LR_predict}a shows the learning curve of the validation set r$^2$ vs epoch for each quantity along with the mean value and Fig.\ref{LR_predict}b-d show the predicted vs. actual values. Interestingly, the ICCD image sequence alone has reasonable predictive power for the growth kinetics, with a mean r$^2_{val}$ = 0.815. Since thin film growth is essentially dominated thermodynamics, it is unexpected to achieve such good predictions without knowledge of the substrate temperature. We believe that the images are correlated to the growth kinetics in this case because the plume intensity (brightness) from each target along with the temporal dynamics (kinetic energy) are directly, but not trivially, correlated to the flux term $J$ and the sticking coefficients $s_0$ and $s_1$, respectively. While $s_0$ and $s_1$ should be strongly correlated to the substrate temperature, the flux ratio (brightness) from the targets will also affect the growth rate and high kinetic energy may cause resputtering\cite{Fahler_1999}. While these factors can be intuitively rationalized, the specific features of the plume dynamics that play a driving role in the growth kinetics are not clear beforehand, which underscores the value of deep learning with plume diagnostics.

By comparison, an MLP trained with only the growth parameters performs slightly better than the ICCD model with r$^2_{val}$ = 0.835. Fig. \ref{LR_predict}e-h show the training results of the MLP. However, the MLP predictions are less stable during training, evident by the large variation in r$^2$ values. The best model captured during training has poor performance on the training set for $s_0$ and $S_1$. While this model accounts for temperature in the growth kinetics, the unstable performance is likely caused by experimental uncertainties related to \textit{E}$_1$ and \textit{E}$_2$. While this MLP has values for the laser energy, the real dynamics of the plume (partially determined the laser) are captured in the ICCD model which may explain why it has more stable predictions. Finally, we train a combined model and the results are shown in Fig. \ref{LR_predict}i-l). The combined model outperforms both previous models with an r$^2_{val}$ = 0.847 with more stable predictions than the MLP. This result shows great promise ML models using combined multimodal PLD diagnostic measurements with basic process parameters for use in autonomous workflows. 

The combined model can be used to predict the growth kinetics in a larger synthesis parameter space than the current dataset explored and identify conditions where the growth times or nucleation rates fall within specified values. Notably, the standard Bayesian optimization (BO) that was used in the growth study that generated this dataset\cite{HarrisAutoPLD} did not incorporate any plume imaging data. In experiments utilizing BO for microscopy, for example, it has been shown that utilizing the local image structure is highly beneficial for optimizing the targeted material properties, which is done by constructing a deep kernel learning (dKL) model \cite{Liu_2022} to map the local image patches to the functional property measured by a spectroscopic measurement. In this case, the equivalent is the multimodal diagnostics data that accompany each deposition. However, in the microscopy case the high-resolution image is acquired prior to the spectroscopy, so all the image patches associated with individual pixels in the scan region are available to train the model at all times. Here, we note that it may be possible to utilize a similar dKL approach by for instance, performing ICCD-only experiments, which can be much faster (and cheaper) than completing an entire film deposition, to acquire this information. More approaches utilizing multi-fidelity Bayesian optimization are also possible, where this model would be considered the low-fidelity approximator, whereas the full film growth would be considered the high-fidelity portion. 

\subsection{Feature map analysis}
\begin{figure*}
\includegraphics[width=\linewidth]{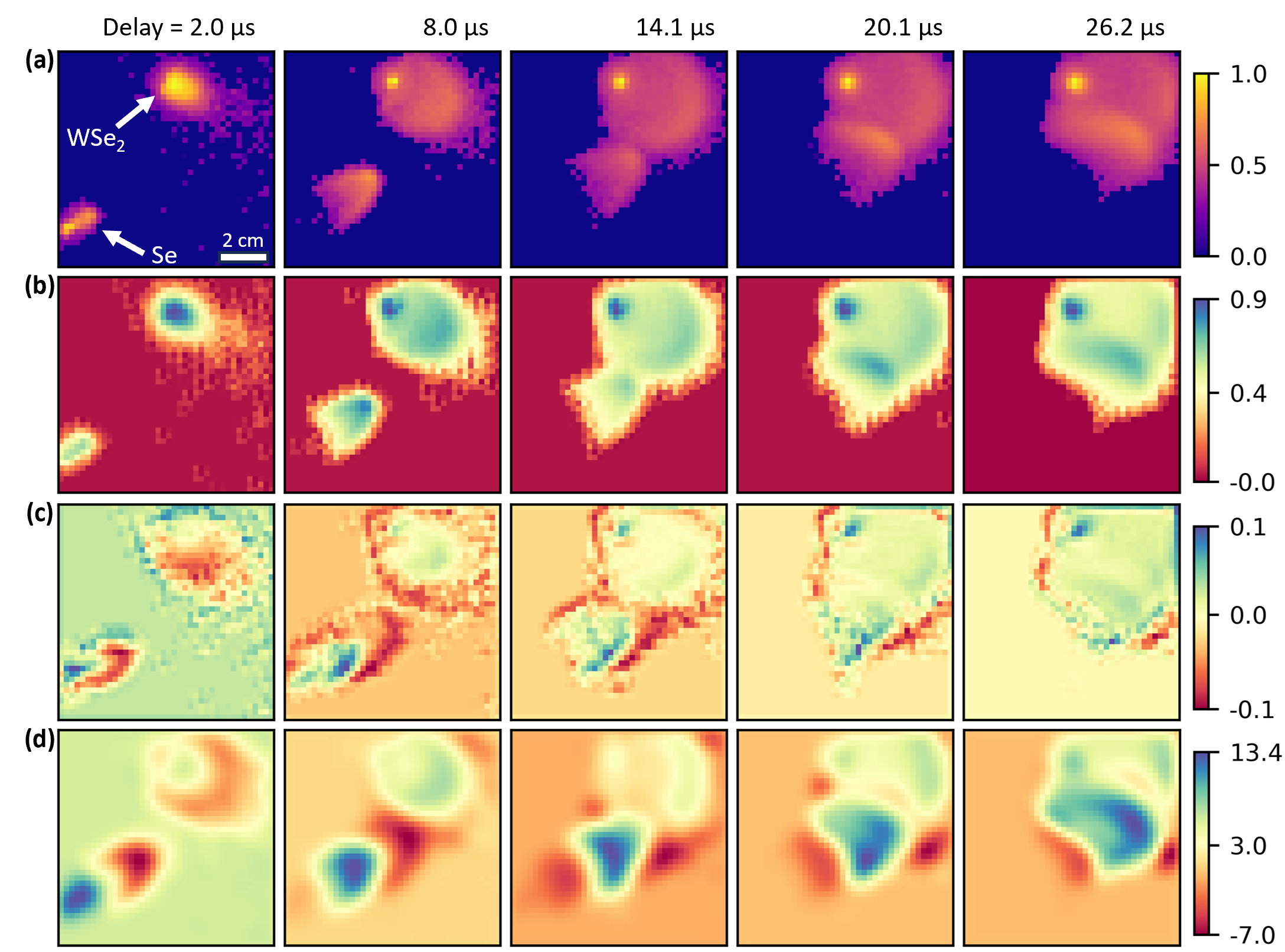}
\caption{\label{fmviz}Activations for select feature maps (FM) of the convolutional layers in the $(2+1)$D convolutional neural network model for growth kinetics highlight the deep features learned from the plumes generated during pulsed laser deposition. (a) Five frames from the processed input image at delay times between 2.0-26.2 $\mu$s show the expansion and collision of the WSe$_2$ and Se plumes. (b) FM from the spatial convolution of layer 1 is similar to the input and discerns the plumes from the background. (c) FM from the 2nd layer spatial convolution appears to function like an edge filter, giving very little weight to the background and plume centers while highlighting the edges. (d) The 3rd layer's temporal convolution encodes spatiotemporal dynamics, giving positive weight to the current location of each plume body and negative weight near edges in the direction of acceleration/deceleration in each frame. }
\end{figure*}

Analyzing the feature maps (FMs) or that are learned by the convolutional layers can potentially provide insight into the features of the plume dynamics that relate to the target. To explore what has been learned by the network, we visualize the activations of the growth kinetic model's $(2+1)$D CNN FMs with an example ICCD image sequence. Many of the FMs are difficult to interpret or look similar with subtle differences between them. Here, we show selected FMs which we can intuitively interpret. Fig. \ref{fmviz}a shows 5 frames from a processed input image with delay times from 2-26.2 $\mu$s. The individual WSe$_2$ and Se plumes can be easily distinguished at the top and bottom of the first frame (2 $\mu$s). As plume expansion progresses, the Se plume collides with the WSe$_2$ plume (14.1 $\mu$s) which induces additional luminescence at the collision fronts (20.1 $\mu$s), likely caused by excitation of neutral atomic species and then the plumes intermix.

Fig. \ref{fmviz}b shows the activations of a FM from the 1st spatial convolution. Many of the initial FMs look similar to the input image which suggest that the early layers are learning simple features such as distinguishing the PLD plumes from the background as well as acting like simple edge filters. In Fig. \ref{fmviz}b, the filter gives no weight to the background and positive values to the plume regions that is correlated with optical emission intensity. Fig \ref{fmviz}c shows a FM from the spatial convolution in layer 2. This FM now includes some temporal dynamics, with positive and negative weights added in each frame in the direction of travel and also tends to highlight the edges of the plumes. The most complex activation we show is from the temporal convolution of layer 3, shown in Fig. \ref{fmviz}d. In this case, positive weight seems to be given to the current location of the plume while negative weight could be interpreted as the direction of plume edge acceleration/deceleration. For instance, at 2 $\mu$s the Se plume is shown as positive weight, corresponding the the plume location in Fig. \ref{fmviz}a. The negative region is indicating the direction of acceleration of the leading edge. At 14.1 $\mu$s, the Se plume has collided with the WSe$_2$ plume and the front begins to accelerate laterally, expanding the width of the plume font. Meanwhile 14.1 $\mu$s, the back edge of the Se plume will begin to decelerate as it "snowplows" into the rest of the mass, indicated by the red region on the back behind the Se plume. The activations of the deepest convolutional layers clearly encode complex spatiotemporal dynamics, which is likely the origin of the increased performance relative to the model trained with a single image (Sec. \ref{anomsec}).

We also visualized the saliency maps using the input gradient method \cite{Simonyan_2014}, which computes pixel importance by multiplying the input image by the gradient of the model output. The saliency maps are also difficult to interpret but provide a method to visualize which regions of the image sequence are important for predictions. Based on the saliency maps, we find that the model is focused on features of the plume, rather than random areas or the background, which provides more support that spatiotemporal dynamics of the plume are critical features for model prediction. Saliency map visualizations are available in the code provided with this article.

\section{Conclusion}
In this work, we investigated the viability of utilizing deep learning with intensified-CCD (ICCD) image sequences of the plasma plume generated during pulsed laser deposition (PLD) for real-time feedback and predictions during thin film growth. Our findings indicate that deep learning, particularly the use of $(2+1)$D convolutional neural networks, can effectively extract complex spatiotemporal features from PLD plume dynamics that are correlated with synthesis conditions and growth kinetics. We demonstrated that ICCD images are highly correlated with chamber pressure and laser energy, thereby providing a means for real-time plume monitoring to detect anomalous conditions during long, unsupervised autonomous experiments. We also showed that plume dynamics are a viable predictor of a thin film growth kinetics model parameters and that when incorporated with basic synthesis parameters, the model performance is increased. These two case studies highlight how the the marriage of in situ plasma diagnostics and machine learning can provide real-time feedback for PLD synthesis with predictive power over the growth environment and materials properties. This work serve as a proof-of-principle for future applications of ICCD images for use in deep kernel learning or multi-fidelity optimization experiments with PLD. We anticipate that this work will encourage more wide-spread adoption of plume imaging techniques, including lower cost CMOS imaging, during PLD for increased reproducibility and accelerated optimization or discovery of materials.
\begin{acknowledgments}
The machine learning in this work was supported by the Center for Nanophase Materials Sciences (CNMS), which is a US Department of Energy, Office of Science User Facility at Oak Ridge National Laboratory. The diagnostics data used in this work was generated with support by the U.S. Department of Energy, Office of Science, Basic Energy Sciences, Materials Sciences and Engineering Division. This research used birthright cloud resources of the Compute and Data Environment for Science (CADES) at the Oak Ridge National Laboratory, which is supported by the Office of Science of the U.S. Department of Energy under Contract No. DE-AC05-00OR22725.

\end{acknowledgments}
\section{Author Declarations}
\subsection{Conflict of Interest}
The authors have no conflicts to disclose.

\subsection{Author Contributions}
\textbf{SBH}: Conceptualization (equal); Writing – original draft (lead); Writing – review \& editing (equal); Data Curation (lead); Methodology (lead); Software (lead); Visualization (lead).
\textbf{CMR}: Resources (equal); Conceptualization (supporting); Writing – review \& editing (equal).
\textbf{KX}: Resources (equal);  Writing – review \& editing (equal).
\textbf{RKV}: Conceptualization (equal); Writing – review \& editing (equal); Software (supporting). All authors read and approved the final manuscript.

\section{Data Availability}
The data that support the findings of this study are openly available at https://github.com/sumner-harris/Deep-Learning-with-ICCD-Images.

\section{Code Availability}
The code for this study can be accessed via this link https://github.com/sumner-harris/Deep-Learning-with-ICCD-Images.
\bibliography{main}
\end{document}